\documentstyle[pre,aps,epsf]{revtex}
\begin{document}
\draft

\twocolumn[\hsize\textwidth\columnwidth\hsize\csname@twocolumnfalse\endcsname

\title{Nucleation, growth, and scaling in slow combustion}
 
\author{Mikko Karttunen$^{1}$, Nikolas Provatas$^{2,3}$,
Tapio Ala-Nissila$^{3,4}$, and Martin Grant$^{1}$}

\address{$^1$ Department of Physics and the Centre for the Physics of
Materials, Rutherford Building, 3600 rue University,\\
Montr\'eal (Qu\'ebec), Canada H3A 2T8}

\address{$^2$Department of Physics and Mechanical Engineering,
University of Illinois at Urbana-Champaign,
Loomis Laboratory of Physics,
1110 West Green Street, Urbana, IL, 61801-3080, U.S.A.}

\address{$^3$Helsinki Institute of Physics,
University of Helsinki, P.O. Box 9, FIN-00014 University of Helsinki, 
Helsinki, Finland}

\address{$^4$Department of Physics, Brown University, Providence, Rhode Island
02912, U.S.A.}

\date{\today}

\maketitle

\begin{abstract}

We study the nucleation and growth of flame fronts in slow combustion.
This is modeled by a set of reaction-diffusion equations for the temperature
field, coupled to a background of reactants and augmented by a term
describing random temperature fluctuations for ignition.
We establish connections between this model and the classical theories
of nucleation and growth of droplets from a metastable phase. Our
results are in good argeement with theoretical predictions.

\end{abstract}

\pacs{PACS numbers: 64.60.My,05.40.+j,82.40.Py,68.10.Gw}


\vskip1pc]
\narrowtext

The kinetic process by which first-order phase transitions take place is an
important subject of longstanding experimental and theoretical interest
\cite{gunton83a}.  Nucleation is the most common of first-order transitions,
and remains of a great deal of interest
\cite{eru95a,metoki90a,kapral94a,yamada84a,duiker90a}.  There are two
fundamentally different cases, homogeneous and heterogeneous nucleation. 
Homogeneous nucleation is an intrinsic process where embryos of a stable
phase emerge from a matrix of a metastable parent phase due to spontaneous
thermodynamic fluctuations.  Droplets larger than a critical size will grow
while smaller ones decay back to the metastable phase
\cite{becker35a,langer69a}.  More commonplace in nature is the process of
heterogeneous nucleation.  There, impurities or inhomogeneities catalyze a
transition by making growth energetically favorable. 

Here we show that the concepts of nucleation and growth can be usefully
applied to understand some aspects of slow combustion.  We use a phase-field
model of two coupled reaction-diffusion equations to study the nucleation
and growth of combustion centers in two-dimensional systems.  Such continuum
reaction-diffusion equations have been used extensively in physics,
chemistry, biology and engineering to describe a wide range of phenomena
from pattern formation to combustion.  However, the connection of
reaction-diffusion equations to nucleation and interface growth has received
little attention. 

In a recent study of slow combustion in disordered media, Provatas {\it et
al.\/}\ \cite{provatas95a,provatas95aa} showed that flame fronts exhibit a
percolation transition, consistent with mean field theory, and that the
kinetic roughening of the reaction front in slow combustion is consistent
with the Kardar-Parisi-Zhang (KPZ) \cite{kardar86a} universality class.  In
this paper we make a further connection between slow combustion started by
spontaneous fluctuations, and the classical theory of the nucleation and
growth of droplets from a metastable phase. 

We generalize the model of Provatas {\it et al.\/}\ by including an
uncorrelated noise source $\eta({\bf x},t)$, as a function of position 
${\bf x}$ and time $t$. The model then consists of equations of motion 
for the temperature field $T({\bf x},t)$
and the local concentration if reactants  $C({\bf x},t)$. The temperature satisfies
\begin{equation}
\frac{\partial T({\bf x},t)}{\partial t} =  
D\nabla^2T-\Gamma[T-T_0]+ 
R(T,C)+\eta.  \label{eq:modelt}
\end{equation}
The first term on the right-hand-side accounts for thermal diffusion,
with diffusion constant $D$, the second term gives Newtonian cooling due to
coupling with a heat bath of background temperature $T_0$, with rate
constant $\Gamma$, and the third term $R(T, C)$ is the exothermic reaction
rate as a function of temperature and concentration of reactants. 
The concentration satisfies \begin{equation} \frac{\partial C({\bf
x},t)} {\partial t} = -\lambda_1 R(T,C), \end{equation} and the reaction
rate obeys \begin{equation} R(T,C)=\lambda_2 T^{3/2}e^{-A/T}C.
\label{eq:modelc} \end{equation} where $\lambda_{1,2}$ are constants, $A$ is
the Arrhenius energy barrier, and Boltzmann's constant has been set to
unity.  The noise is assumed to be uncorrelated and Gaussian of zero mean
with a second moment $\langle \eta({\bf x},t)\eta({\bf x}',t') \rangle =
2\epsilon \delta({\bf x}-{\bf x}') \delta(t-t')$, where the angular brackets
denote an average, and $\epsilon$ is the intensity of the 
noise. Note that while the dynamics of the process is controlled by the
activation term $e^{-A/T}$, the scale for energy is set by $T^{3/2}$. We
choose the same values for the constants as those used in
Ref.~\cite{provatas95aa}, which are approximately those for the combustion
of wood in air: In physical units $D = 1$ m$^2$s$^{-1}$, $\Gamma = 0.05$
s$^{-1}$, $T_0 = 0.1$ K, $A = 500$ K, and the specific heat of wood, $c_p =
5$ Jg$^{-1}$K$^{-1}$ (entering through $\lambda_2 = 8$).  Length is measured
in units of the reactant size, and time in units of those for the reaction
to take place, $\lambda_2/\lambda_1$. The low background temperature is
chosen to permit good numerical accuracy for systems of moderate size. The
parameter $\epsilon$ was varied in the range $2 \times 10^{-7}-11\times
10^{-7}$, and models the spontaneous fluctuations in heat in a random
medium: $\epsilon=2 \times 10^{-7}$ corresponds to slow, $\epsilon=5\times
10^{-7}$ to medium, and $\epsilon=11 \times 10^{-7}$ to fast nucleation rate. 

\begin{figure}[hbt]
\epsfxsize=\columnwidth\epsfbox{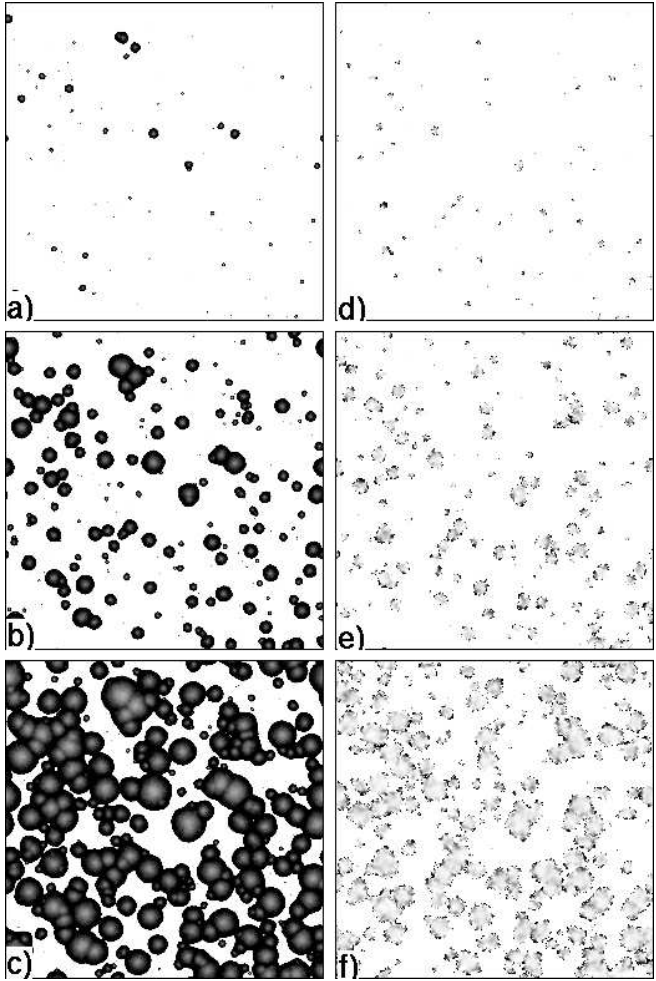}
\vspace{15pt}
\caption{A snapshot of the process with medium nucleation rate, where
$\epsilon = 4.9 \times 10^{-7}$. 
Gray scale images temperature, where black is the hottest region.
The left column shows a
system with uniform background ($c=1$) and the right column 
a case where $c=0.5$.
In Fig.\ 1 a)-c) the time steps are $t=8$, $t=12$ and 
$t=16$, and in Fig.\ 1 d)-f)  $t=10$, $t=15$ and $t=20$.
The right panel features burning domains at significantly lower 
temperature, and with more ragged boundaries, due to the lower 
concentration of reactants.}
\label{fig:snap}
\end{figure}

We integrate these equations using Euler difference rules in space and time,
with the smallest length $\Delta x = 1$, and time $\Delta t =0.01$. Reactant
units are randomly dispersed across the grid points with a probability $c$
so that $C({\bf x},t)=0$ for an unoccupied site and $C({\bf x},t)=1$ for an
occupied site. Figure~\ref{fig:snap} shows typical results for a
two-dimensional system of size $256 \times 256$, with periodic boundary
conditions.  The gray scale images different temperatures, with black being
the hottest regions. The left-hand panel shows the case for uniform
concentration $c = 1$, while the right-hand panel shows cooler and more
ragged flame fronts that occur for $c = 0.5$ (which is still well within the
concentration for which the flame front can propagate, i.e. where $c \ge
c^*$, and $c^* \approx 0.2$ for $\epsilon = 0$). 

The morphology of burnt and unburnt zones in these figures is strikingly
similar to that for the nucleation and growth of crystallites from a
supercooled melt \cite{gunton83a}, which is the motivation for our approach. 
We now briefly review the classical theories of nucleation and growth for
such systems where a conservation law does not control the growth process. 

The classical theory of nucleation was formulated in the 1930's by Becker
and D\"oring \cite{becker35a}. Their phenomenological theory has two main
results:  A description of the critical droplet, based on free energy
considerations, and a rate equation for the growth of clusters. There exists
also a modern theory \cite{langer69a}, derived from first-principles, which
generalizes the classical theory by, e.g., taking into account the finite
interface thickness of the droplet domain walls. However, within the scope
of this work, it is adequate to consider only the classical theory. In the
classical theory, the extra free energy due to a droplet of stable phase, in
a metastable background, is $\Delta F = - V \Delta f + A \sigma$, where $V$
is the volume of the droplet, $A$ is its area, $\sigma$ is surface tension,
and $\Delta f$ is the difference in the bulk free energy densities between
the metastable and stable phases.  The critical radius $\rho^*$ of a droplet
is obtained from through minimization: ${\partial \Delta F} / {\partial
\rho^*} = 0$. For circular droplets in two dimensions, this gives
$\rho^*={\sigma}/ {\Delta f}$ for the critical radius, and $\Delta
F(\rho^*)={\pi \sigma^2} / {\Delta f}$ for the height of the free energy
barrier.  For the metastable phase to decay, droplets with $\rho > \rho^*$
nucleate and grow; droplets smaller than the critical radius shrink and
disappear.  The rate-limiting process is the critical droplet, with energy
barrier $\Delta F(\rho^*)$, whose probability of occurrence is proportional
to $\exp[-\Delta F(\rho^*)/T]$. 

After nucleation has occurred, the subsequent growth of droplets is often
well described by the phenomenological Kolmogorov-Avrami-Mehl-Johnson (KAMJ)
\cite{kolmogorov37a,avrami39a,johnson39a} model. This treatment describes
many solid-solid and liquid-solid transformations, provided long-range
interactions between droplets (which can be due to elastic effects, or
diffusion fields) are of negligible importance.  The KAMJ theory assumes
that nucleation is a non-correlated random process with isotropic droplet
growth occurring at constant velocity, where the critical radius is
infinitesimal, and growth ceases when growing droplets impinge upon each
other.  As its basic result, the KAMJ description gives a functional form
for the volume fraction of the transformed material: 
\begin{equation}
X(t)=1- \exp [-\frac{V}{d+1}Iv^d(t-t_0)^{d+1}]
\label{eq:avrami1}
\end{equation}
for homogeneous nucleation, and
\begin{equation}
X(t)=1- \exp [-V \alpha v^d(t-t_0)^d]
\label{eq:avrami2}
\end{equation}
for heterogeneous nucleation.  In the above, $v$ is the growth velocity, $I$
is the nucleation rate and $\alpha$ is the density of embryos with $\rho >
\rho^*$ present in the beginning of the process. Homogeneous and
heterogeneous nucleation are conveniently distinguished by the Avrami
exponent, which is $d+1$, in Eq.~(\ref{eq:avrami1}), and $d$ in
Eq.~(\ref{eq:avrami2}). The waiting time $t_0$ in Eqs.~(\ref{eq:avrami1})
and (\ref{eq:avrami2}) accounts for initial transients. 

In the KAMJ description there are two intrinsic length scales present in the
system. The first one is the critical radius and the other is found by
simple dimensional analysis. As seen from Eqs.~(\ref{eq:avrami1}) and
(\ref{eq:avrami2}), the process is characterized by two variables: the
nucleation rate and the growth velocity. Using them, the characteristic
length can be written as $\xi = (v/I)^{1/(d+1)}$ and the characteristic time
scale as $\tau = (Iv^d)^{-1/(d+1)}$. In practice, it is convenient to scale
the time by the half-time of the transformation, $t_{1/2}$, since it is an
easily accessible quantity both experimentally and computationally, and it
can be used as a measure \cite{yamada84a}.  In the limit where $\xi \gg
\rho^*$ there is only one length scale present \cite{ishibashi71a}.
During nucleation and growth
this is the case up to the point when a connected cluster spans the system. 
As a consequence of this, scaling of $X(t)$ is expected, as we will show
below. 

The apparent simplicity of the KAMJ description is due to the fact that it
incorporates no correlations.  For cases where such correlations are minimal
(as for the case considered herein), it has been quite successful in
describing experimental data \cite{metoki90a,yamada84a}, and theoretical
generalizations can be readily made
\cite{eru95a,duiker90a,bradley89a,ishibashi71a}. The KAMJ description can be
used in calculations of kinetic parameters and activation energies, and it
provides information about the nature of the phase transition, i.e. if the
process is diffusion or reaction (interface) controlled and if the process
is influenced by inhomogeneities.  Unfortunately, the basic KAMJ theory
provides no information about the structural changes occurring during the
phase transformation.  Based on the same assumptions, Sekimoto
\cite{sekimoto86a} derived exact analytical expressions for two-phase
correlation functions when $\xi \gg \rho^*$.  Fourier transforming
Sekimoto's result for the two-point equal-time correlation function gives
the structure factor: 
\begin{equation}
S({\bf k},t)=\int d{\bf r}\mbox{~}[C_2({\bf x},t;{\bf x}+{\bf
r},t)-C_1^2({\bf x},t)]e^{i{\bf k}\cdot {\bf r}},
\label{eq:structure}
\end{equation}
where $C_1({\bf x},t)$ 
is the one-point correlation function equal to the KAMJ expression 
for transformed
volume given in Eqs.~(\ref{eq:avrami1}) and (\ref{eq:avrami2}). The
two-point correlation function $C_2$ is
\begin{equation}
C_2({\bf x},t;{\bf x}+{\bf r},t)= C_1^2({\bf x},t)\exp [Iv^2\Psi(y)],
\label{eq:c2}
\end{equation}
where 
\begin{eqnarray}
\Psi (y) & = & \frac{2}{3}\left[ \arccos (y) -2y\sqrt{1-y^2} \right. \nonumber \\
& & + y^3 \left. \ln \left( \frac{1+\sqrt{1-y^2}}{y}\right) \right ],
\label{eq:psi}
\end{eqnarray}
for $y \le 1$, and $\Psi(y) = 0$ for $y>1$. The variable 
$y$ is the normalized distance between two points.
Combining Eqs.~(\ref{eq:structure}), (\ref{eq:c2}), and (\ref{eq:psi})
gives 
\begin{eqnarray}
S({\bf k},t)& = & 2 \pi a^2e^{-\frac{2}{3}\pi I v^2 t'^3} \times \nonumber \\
& & \int_0^1 dy\mbox{~}[e^{ Iv^2t'^3 \Psi (y)}-1]yJ_0(aky),
\label{eq:stu}
\end{eqnarray}
where $J_0$ is the Bessel function of the zeroth kind, and 
$a=2vt'$  with $t'=t-t_0$. 

We expect these theories to give a reasonable description of the growth
of the flame fronts.  In fact the agreement is much more impressive
than we had anticipated.   Of course, for combustion, the picture of
nucleation and growth must be modified or re-interpreted in
straightforward ways.  For example, no shrinking of droplets, which
herein correspond to burned patches, is possible.  Also, instead
of temperature in the Boltzmann probability weight, a quantity
proportional to the intensity of noise sources $\epsilon$ must appear.
Furthermore, the surface tension, evident in the fact that the burned
patches are round, must have its origin in the dynamical description.
Indeed, in the absence of nucleation it has been shown in
Ref.~\cite{provatas95aa} that the flame front roughens according to
the KPZ interface equation, through which some of these correspondences
can be made.  

For example, the critical radius can be found as follows.
We follow the method of Ref.~\cite{kapral94a,hamalainen96a}, and write
the KPZ equation in circular coordinates as
\begin{equation} 
\frac{\partial \rho}{\partial t}=
  \frac{D}{\rho^2} \frac{\partial^2 \rho}{\partial \theta^2}
-  \frac{D}{\rho}
+  v\left(1+\frac{\rho^2_{\theta}}{2\rho^2}\right) +
     \frac{1}{\sqrt{\rho}}\mu(\theta,t), \label{eq:rkpz} \end{equation}
where $\mu$ is a linear combination of an effective noise due to the random
reactant, and the additive noise $\eta$. Next, we use the positive part of
Eq.~(46) in Ref.~\cite{provatas95aa} with $v=(\Gamma \Lambda - \hat{\lambda}
c)/\sigma$. The constants $\Lambda$, $\hat{\lambda}$ and $\sigma$ depend
functionally on the temperature $T_m(x)$ that solves Eqs.~(1) and
(\ref{eq:modelc}) in the mean field limit. Their exact forms are given in
Ref.~\cite{provatas95aa}.  Physically, the constant $\Lambda$ is
proportional to the heat loss in the mean field limit, $\hat{\lambda}$ is
proportional to the heat produced at the interface in the mean field limit,
and $\sigma$ is analogous to surface tension.  To find an expression for the
critical radius, we apply perturbative analysis. Expanding $\rho$ and $\mu$ as
$\rho(\theta,t)=\sum_n \rho_n(t) e^{in \theta}$, and $\mu(\theta,t)=\sum_n
\mu(t) e^{in\theta}$, and substituting them into Eq.~(\ref{eq:rkpz})
together with the velocity gives $\rho^*=D\sigma/(\Gamma \Lambda -
\hat{\lambda} c)$ for the the lowest order estimate for the critical radius
of a radial flame front.  When $\Gamma \Lambda \rightarrow \hat{\lambda} c$,
$v\rightarrow 0$. 
In this limit the flame front does not propagate, since the heat lost to
thermal dissipation exactly balances that due to thermal reaction, and the
critical radius goes to infinity.  Unfortunately, the numerical window in
which the critical radius changes appreciably is narrow, and close to $c^*$.
Hence, although our numerical work reported below is consistent with the
above analysis, it does not permit a quantitative test of our estimate of
$\rho^*$. 

\begin{figure}[hbt]
\epsfxsize=\columnwidth\epsfbox{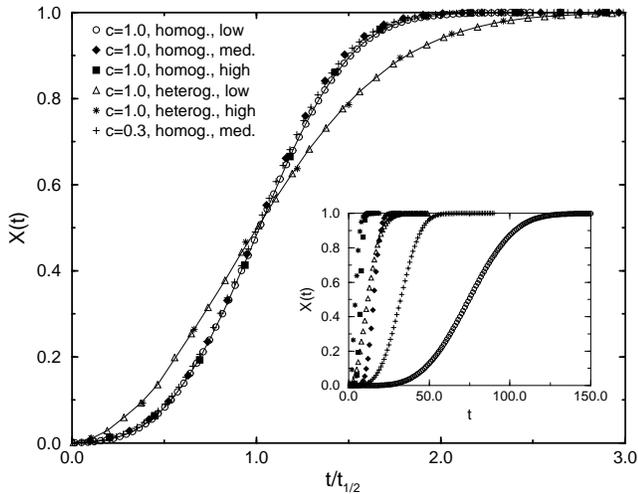} 
\caption{Fraction burned vs.
$t/t_{1/2}$ for various nucleation rates in homogeneous and heterogeneous
nucleation in uniform and disordered systems.  The inset shows the same data
sets without scaling by $t_{1/2}$. The data is indistinguishable from the
KAMJ theory (solid lines).} 
\label{fig:kajm} 
\end{figure}

In our simulations we have typically used lattices of size $256 \times 256$
with periodic boundary conditions, averaging over 1000 sets of initial
conditions. Heterogeneous nucleation is modeled via inital fluctuations 
at $t=0$, and homogeneous nucleation as time dependent Gaussian 
fluctuations through $\eta$.
First, we compare our results for the fraction of burnt reactant
product $X(t)$ to the KAMJ theory. We have simulated systems with various
noise intensities with uniform and disordered reactant concentrations to see
the applicability of the KAMJ theory in relation to this model. In both
uniform and disordered cases the simulations are in excellent agreement with
the theory as seen in Fig.~\ref{fig:kajm}.  
The case of homogeneous nucleation fits the Avrami exponent $3$, while the
heterogeneous case gives $2$, as expected.
In the scaled plots we have discarded the waiting time $t_0$
(Eqs.~(\ref{eq:avrami1}) and (\ref{eq:avrami2})) since it is due to lattice
effects and the fairly small system size, and therefore does not represent a
true physical time scale. 

\begin{figure}[hbt]
\epsfxsize=\columnwidth\epsfbox{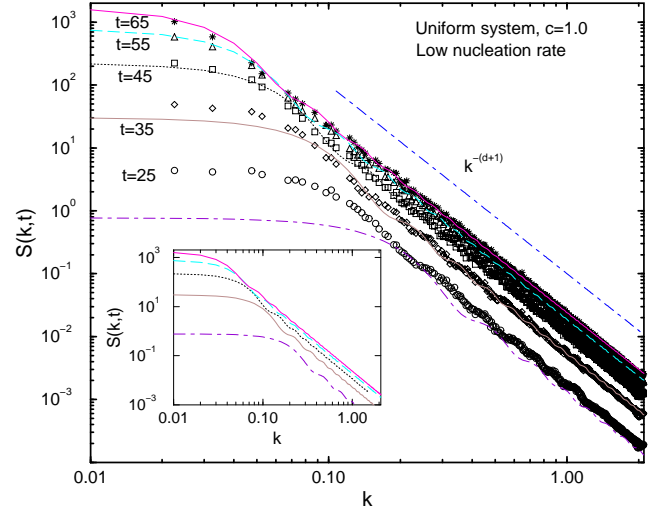}
\caption{A plot of the structure factor vs. $k$ for various time steps
at a low noise intensity (see Fig.~{\protect \ref{fig:kajm}}) for
a uniform system.  The long-dashed line shows Porod's law, which 
describes large-$k$ correlations of randomly-oriented interfaces of
negligible width. In the
inset we have replotted the theoretical curves to demonstrate the
presence of wiggles more clearly.}  
\label{fig:strk1}
\end{figure}

\begin{figure}[hbt]
\epsfxsize=\columnwidth\epsfbox{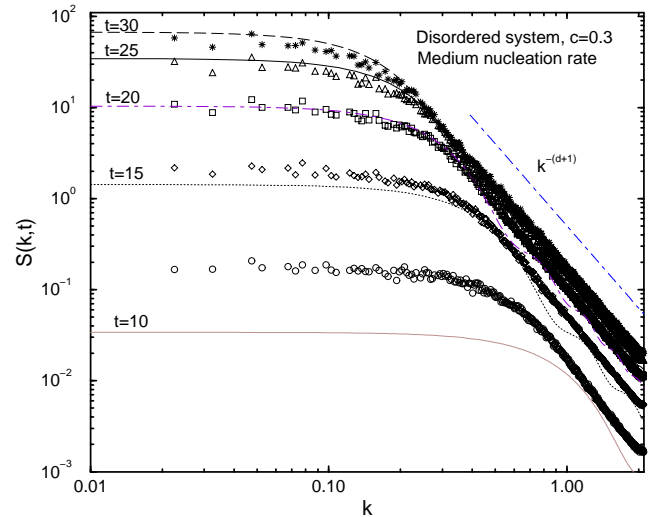}
\caption{A plot of the structure factor vs. $k$ for various time steps
at medium nucleation rate (see Fig.~{\protect \ref{fig:kajm}}) for 
a disordered system. 
The dot-dashed line shows Porod's law. }  
\label{fig:strk2}
\end{figure}

Next, we will focus on the structure factor in the case of uniform
homogeneous nucleation, and compare $S({\bf k},t)$ for the temperature field
from the simulations to Sekimoto's theoretical prediction,
Eq.~(\ref{eq:stu}), for various cases.  Here, $S({\bf k},t)$ corresponds to
correlations in reactant concentrations, i.e., between burnt zones.  In the
cases of both high and low noise, we find a quantitative agreement between
the theory and simulations at late times, as seen
Figs.~\ref{fig:strk1} and \ref{fig:strk2}. In order to use the theoretical
prediction, Eq.~(\ref{eq:stu}), we measured the growth velocity of the
radius of individual nucleation centers for various concentrations, and
found it to be in agreement with previous results
\cite{provatas95a,provatas95aa}, i.e. $R(t) \sim t$.  

As seen from Figs.~\ref{fig:strk1} and \ref{fig:strk2}, the theoretical 
prediction underestimates the rate of phase transformation during 
\noindent the early stages, but is in good agreement at later times. 
This is because, for early times, contributions to the structure factor from 
the bulk interior of droplets, and from the diffuse surface width of droplets 
are comparable. For later times, the surface contributions (not considered 
in the KAMJ and Sekimoto theories) are negligible.

\begin{figure}[hbt]
\epsfxsize=\columnwidth\epsfbox{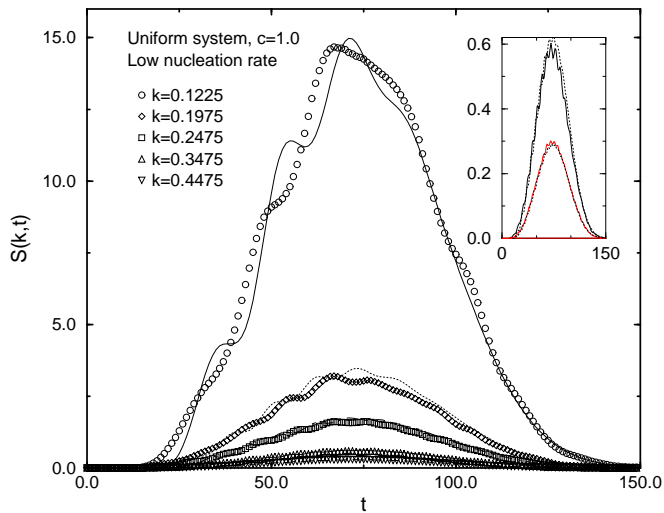}
\caption{The structure factor vs. time for various $k$
at a low noise intensity for a uniform system. 
The symbols represent the data from the
simulations and the lines display the theoretical prediction obtained
by integrating Eq.~(\protect{\ref{eq:stu}}). The inset shows the data
for $k=0.3475$ and $k=0.4475$. The symbols display the data from
the simulations, and the solid lines the theoretical prediction.
}
\label{fig:strt1}
\end{figure}

For a uniform background there are very pronounced wiggles present, as can
be seen from Figs.~\ref{fig:strk1} and \ref{fig:strt1}. Their origin can be
traced to the presence of the Bessel function in Eq.~(\ref{eq:stu}).  The
oscillations are due to the spherical shape of the burnt zones at early
stages of the process 
\begin{figure}[hbt]
\epsfxsize=\columnwidth\epsfbox{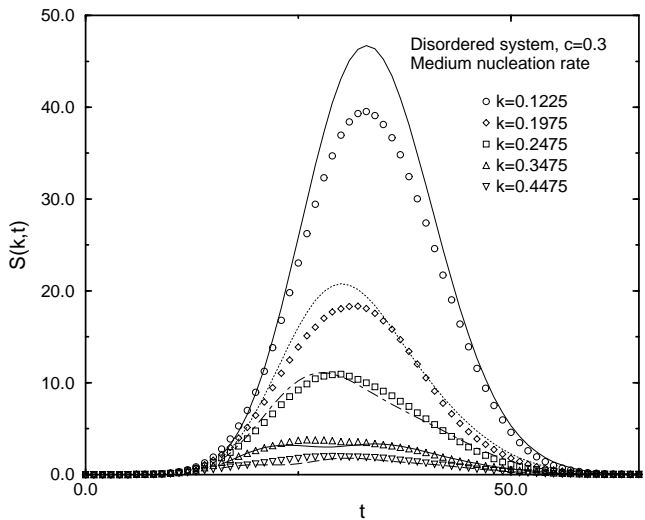}
\caption{
The structure factor vs. time for various $k$
at a medium nucleation rate for a disordered system. 
The symbols represent the data from the
simulations and the lines display the theoretical prediction obtained
by integrating Eq.~(\protect{\ref{eq:stu}}).  } \label{fig:strt2}
\end{figure}
\noindent 
when the growing regions have not yet merged with each
other. That is, for early times, the structure factor is essentially the
structure factor for a single droplet, with radius equal to the mean. The
absence of wiggles in the case of quenched disorder, Fig.~\ref{fig:strk2}
and Fig.~\ref{fig:strt2}, is simply due to the fact that the disorder
affects the spreading of the temperature field resulting in kinetic
roughening.  This is also clearly visible in Fig.~\ref{fig:snap}. In
Figs.~\ref{fig:strk1} and ~\ref{fig:strk2} we have also compared the
numerical results to Porod's law, $S(k,t) \sim 1/k^{d+1}$ at large $k$
\cite{gunton83a}. This should describe a locally flat and thin interface, 
and indeed we find excellent agreement for large $k$.

To confirm that the origin of the discrepancies between the theory and our
simulations of the structure factor are due to the interface width, we
also compared the results from the combustion model to a two-dimensional
cellular automaton model with a nearest neighbor updating rule, using
lattices of size $256 \times 256$ and $1024 \times 1024$. Using strictly
nearest neighbor interactions for linear growth, so that there was no
disorder in the front region, the CA model matched exactly the KAMJ result
for the volume fraction (Eq.~(\ref{eq:avrami1})), and Sekimoto's result for
the structure factor (Eq.~(\ref{eq:stu})), as expected.

The constant growth velocity, and the scaling of $X(t)$ might suggest that
the structure factor exhibits scaling, with the characteristic length
increasing linearly in time, $L(t) \sim t$.  However, as is evident from
Figs.~\ref{fig:strk1} and \ref{fig:strk2}, this turns out not to be the case.
This can also be seen from Eq.~(\ref{eq:stu}). At early times the structure
factor follows approximately  $ \sim t^5$, and for late times it falls off
exponentially.  The reason is that the KAMJ theory applies to uncorrelated
systems.  The constant-velocity growth of a single domain is essentially due
to the constant driving force of an excess chemical potential. But the
distribution in sizes and in space of the droplets is due to their time and
position of nucleation, implied by the nucleation rate.  These two time
scales are not proportional to each other, so no scaling results.

To conclude, we have studied the connection between the classical theory of
nucleation and growth, and a model of slow combustion. We find that the
reaction occurs with constant disorder dependent velocity with a linear
scaling for the characteristic length $L(t)$ for the individual growing
clusters. We have studied the structure factor of the temperature field and
found good agreement with the theoretical predictions \cite{sekimoto86a}. 
These results could be tested in a two-dimensional reaction-diffusion cell,
or simply by slowly burning uncorrelated paper, i.e., with insignificant
convection. 

\acknowledgements
We would like to thank Ken Elder for helpful conversations.  This work has
been supported by the Academy of Finland, The Finnish Cultural Foundation,
the Finnish Academy of Science and Letters, the Natural Sciences and
Engineering Council of Canada, and {\it les Fonds pour la Formation de
Chercheurs et l'Aide \'a la Recherche de Qu\'ebec\/}. In addition, we wish
to thank the Centre for Scientific Computing (Espoo, Finland), which has
provided most of the computing resources for this work.


\begin{references}

\bibitem{gunton83a} For a review see J. D. Gunton, M. San Miguel, and
P. S. Sahni, in {\it Phase Transitions and Critical Phenomena}, edited
by C. Domb and J. L. Lebowitz (Academic Press, London, 1983), Vol.~8.

\bibitem{eru95a}    V. Erukhimovitch, J. Baram, Phys. Rev. B {\bf 51},
	            6221 (1995).
\bibitem{metoki90a} N. Metoki, H. Suematsu, Y. Murakami, Y. Ohishi,
		    Y. Fujii,
		    Phys. Rev. Lett. {\bf 64}, 657 (1990).
\bibitem{kapral94a} R. Kapral, R. Livi, G.-L. Oppo, A. Politi,
		    Phys. Rev. E {\bf 49}, 2009 (1994).	 
\bibitem{yamada84a} Y. Yamada, N. Hamaya, J.D. Axe, S.M. Shapiro,
	            Phys. Rev. Lett. {\bf 53}, 1665 (1984).
\bibitem{duiker90a} H.M. Duiker, P.D. Beale, Phys. Rev. B {\bf 41},
                    490 (1990).
\bibitem{becker35a} R. Becker, W. D\"oring, Ann. Phys. (Leipzig) 
	            {\bf 24}, 719 (1935). The classical nucleation
                    theory is reviewed at length in F.F. Abraham, 
	            {\it Homogeneous Nucleation Theory} Academic
                    Press, NY, 1974).
\bibitem{langer69a} J. S. Langer, Ann. Phys. {\bf 54}, 258 (1969).
\bibitem{provatas95a}  N. Provatas, T. Ala-Nissila, L. Pich\'e, M. Grant, 
                       Phys. Rev. E {\bf 51}, 4232 (1995).
\bibitem{provatas95aa}N. Provatas, T. Ala-Nissila, L. Pich\'e, M. Grant, 
                       J. Stat. Phys. {\bf 81}, 737 (1995).
\bibitem{kardar86a} M. Kardar, G. Parisi, and Y. C. Zhang, 
                    Phys. Rev. Lett. {\bf 56}, 889 (1986).
\bibitem{kolmogorov37a}  A.N. Kolmogorov, Bull. Acad. Sci. USSR,
	                 Mat. Ser. {\bf 1}, 335 (1937).
\bibitem{avrami39a} M. Avrami, J. Chem. Phys. {\bf 7}, 1103 (1939).
\bibitem{johnson39a} W.A. Johnson, A. Mehl, Trans. Am. Inst. Min.
                     Eng.  {\bf 135} 416, (1939).
\bibitem{bradley89a} R. M. Bradley, P. N. Strenski,
	             Phys. Rev. B {\bf 40}, 8967 (1989). 
\bibitem{ishibashi71a} Y. Ishibashi, Y. Takagi,
		       J. Phys. Soc. Jpn. {\bf 31}, 506 (1971).
\bibitem{sekimoto86a}  K. Sekimoto, Physica A {\bf 135}, 328 (1986).
	             The exponent of $y$ in front of the logarithmic term should
	             be 3 instead of 2 in Eq.~(3.14) of this paper.
\bibitem{hamalainen96a} It is also possible to derive the radial KPZ equation
	directly for the flame fronts: J. H\"am\"al\"ainen, N. Provatas, T. Ala-Nissila, 
	J. Timonen, unpublished (1996).
\bibitem{karttunen96} M. Karttunen and M. Grant, unpublished (1996).

	
\end{references}
\end{document}